\documentclass[twocolumn,twoside]{ncnsd} 

\def\BibTeX{{\rm B\kern-.05em{\sc i\kern-.025em b}\kern-.08em
    T\kern-.1667em\lower.7ex\hbox{E}\kern-.125emX}}

\begin{document}
\newcommand{\newc}{\newcommand}
\newc{\beq}{\begin{equation}}
\newc{\eeq}{\end{equation}}
\newc{\kt}{\rangle}
\newc{\br}{\langle}
\newc{\beqa}{\begin{eqnarray}}
\newc{\eeqa}{\end{eqnarray}}
\newc{\pr}{\prime}
\newc{\longra}{\longrightarrow}
\newc{\ot}{\otimes}
\newc{\rarrow}{\rightarrow}
\newc{\h}{\hat}
\newc{\bom}{\boldmath}
\newc{\btd}{\bigtriangledown}
\newc{\al}{\alpha}
\newc{\be}{\beta}
\newc{\ld}{\lambda}
\newc{\sg}{\sigma}
\newc{\p}{\psi}
\newc{\eps}{\epsilon}
\newc{\om}{\omega}
\newc{\mb}{\mbox}
\newc{\tm}{\times}
\newc{\hu}{\hat{u}}
\newc{\hv}{\hat{v}}
\newc{\hk}{\hat{K}}
\newc{\ra}{\rightarrow}
\newc{\non}{\nonumber}
\newc{\ul}{\underline}
\newc{\hs}{\hspace}
\newc{\longla}{\longleftarrow}
\newc{\ts}{\textstyle}
\newc{\f}{\frac}
\newc{\df}{\dfrac}
\newc{\ovl}{\overline}
\newc{\bc}{\begin{center}}
\newc{\ec}{\end{center}}
\newc{\dg}{\dagger}
\newc{\prh}{\mbox{PR}_H}
\newc{\prq}{\mbox{PR}_q}

\title{The Fourier transform of the Hadamard transform: Multifractals, Sequences and Quantum Chaos}

\author{N. Meenakshisundaram and Arul Lakshminarayan
\thanks{The authors are with the Department of Physics at the Indian Institute of
Technology Madras. e-mail: arul@iitm.ac.in. It is a pleasure for AL to thank
S. Nonnenmacher, M. Saraceno and A. Voros,for discussions.
NM was financially supported by a CSIR fellowship.}}
\maketitle

\begin{abstract}
We introduce a class of functions that limit to multifractal measures and which arise when 
one takes the Fourier transform of the Hadamard transform. This introduces generalizations
of the Fourier transform of the well-studied and ubiquitous Thue-Morse sequence, and
 introduces also generalizations to other intriguing sequences. We show their relevance to
 quantum chaos, by displaying quantum eigenfunctions of the quantum bakers map that 
 are approximated well by such measures, thereby extending our recent work where we pointed
 to the existence of ``Thue-Morse'' states. 
\end{abstract}

\begin{keywords}
Fourier transform. Hadamard transform. Multifractals. Sequences. Quantum chaos. Eigenfunctions.
\end{keywords}

\section{Introduction}
\PARstart{T}{he} Fourier and Hadamard transforms are standard tools, widely used in 
science and signal processing \cite{Schroeder}. The relative importance of the two transforms may
be judged to be a factor of thirty in favour of the Fourier transform if one were
to go by a "google" search which returned over five million webpages for this transform.
Both these transforms can be implemented with fast algorithms that reduce their
implementation on $N$ data points from $N^2$ to $N\, \log(N)$ operations. The fast 
Fourier transform (FFT) and the fast Hadamard transform essentially rely on the 
factoring of the transform into operators acting on product vector spaces. The Hadamard
transform though is a real transform which only adds or subtracts the data and is therefore 
widely used in digital signal processing. The Fourier transform conjugate spaces are 
familiar ones ("time-frequency", "position-momentum") etc., while the corresponding
Hadamard transforms are not so well understood. Nevertheless the Hadamard transform
has also received great attention in the recent past due to its uses in quantum
computing, with the Hadamard gate being a central construct \cite{Nielsen}.

We define the Fourier transform (FT) on $N$ sites as 
\begin{equation}
(G_N)_{m,n}=\frac{1}{\sqrt{N}}\exp\left(-2 \pi i (m+\alpha)(n+\alpha)/N\right)
\end{equation}
with $0\le \alpha \le 1/2$ being a phase parameterizing the transform, and
$0\le m,n\le N-1$. $G_N$ as a matrix is an unitary one. The Hadamard transform
that we use maybe written in several ways, firstly as an tensor or Kronecker product
form, secondly via a recursion and finally via their matrix elements. In all of this
and what follows in this paper {\it we assume} $N$ {\it to be a power of 2}, i.e., $N=2^K$,
for some integer $K$. If 
\begin{equation}
H_{2}=\frac{1}{\sqrt{2}}\left( \begin{array}{rr} 1 & 1\\ 1 & -1 \end{array} \right) 
\end{equation}
then 
\beq
H_{2^K}=H_2 \otimes H_2 \otimes \cdots \otimes H_2=\otimes^K H_2.
\eeq
Equivalently
\beq
H_{2^{i+1}}=\f{1}{\sqrt{2}}\left( \begin{array}{rr} H_{2^{i}} & H_{2^{i}} \\ H_{2^{i}} & -H_{2^{i}} \end{array} \right), 
\eeq
and $H_{2^0}=1$.
Also in terms of matrix elements
\beq
(H_{N})_{m,n}=\f{1}{\sqrt{N}} (-1)^{a\cdot b}
\eeq
where $a\cdot b=\sum_{i=1}^{K} a_i\, b_i$ and $m=\sum_{i=1}^{K}{a_i}2^{i-1}, \; 
n=\sum_{i=1}^{K}{b_i}2^{i-1},$ that is $a$ and $b$ are vectors whose entries
are the binary expansions of the matrix positions $(m,n)$.
Note that $H_N$ is such that $H_N^2=I$, while $G_N^4=I$, where $I$ is the identity,
therefore the spectrum of both these transforms are highly degenerate ($\pm 1$ for
$H_N$, $\pm 1, \pm i$ for $G_N$). 

Also notice that we can enumerate the columns of $H_N$ (or rows, as it is a symmetric matrix)
as outer products of 
\beq v_0\, =\, \f{1}{\sqrt{2}} \left(\begin{array}{rr}1\\1 \end{array} \right), \;\;
v_1\, =\, \f{1}{\sqrt{2}} \left(\begin{array}{rr}1\\-1 \end{array} \right) \eeq
If $n=a_{K-1}a_{K-2}\cdots a_{0}$ is its binary representation ($0\le n \le 2^K-1$),
the $n^{\mbox{th}}$ column $V_n$ is the outer product
\beq V_n = v_{a_{K-1}}\otimes v_{a_{K-2}} \otimes \cdots \otimes v_0. \eeq
While the first column $V_0$ is simply an uniform string of 1s, the last
one $V_{N-1}$ is the $K$-th generation of the Thue-Morse sequence.

The Thue-Morse sequence is an example of an ``automatic sequence'' \cite{Allpaper}
of two alphabets say $A$ and $B$. Given say $A$, the rule is to replace it 
by $AB$ and given $B$ replace it by $BA$. Thus starting with the seed $A$ we get 
$\{ A \rightarrow AB \rightarrow ABBA \rightarrow ABBABAAB \rightarrow \cdots \}$.
The $K$-th generation consists of a string or word of length $2^K$ which is cube free,
 that is no block (any finite string consisting of the two alphabets) repeats thrice consecutively. This sequence occurs in numerous
  contexts \cite{Allpaper}, combinatorics on words, number theory, group theory, dynamical
systems, to name a few, and is considered to be marginal between a quasiperiodic sequence and a chaotic one. The deterministic
disorder of this sequence is relevant to models of quasicrystals \cite{Bovier}, 
mesoscopic disordered system \cite{Janssen}, and as we established recently to quantum chaos
 \cite{MLpre}.
The column $V_{N-1}$ is got from the $K$-th generation of the Thue-Morse sequence
via the identification $A \equiv 1$ and $B \equiv -1$, apart from the factor $1/\sqrt{N}$.

Our recent work indicated that the Fourier transform of the Thue-Morse sequence,
along with the sequence itself was an excellent ansatz for a class of eigenstates
in the quantum baker map, which we called the ``Thue-Morse states''.
 The classical baker map is a paradigmatic and simple
model of complete Hamiltonian chaos. Its quantization is then of considerable interest
in the study of quantum chaos, therefore the emergence of this ansatz provides
an interesting way to think of the deterministic structural disorder of quantum
chaotic eigenfunctions, at least in this model system. We also found that the Fourier
transform of some other columns of the Hadamard transform played a crucial role
in describing other states. The Fourier transform of the Thue-Morse sequence \cite{Luck,Rao}, or
some of the other columns of $H_N$ are not simple functions though, they could
be multifractals \cite{Halsey}. Of course the Fourier transform of the first column $V_0$ is 
just a localized delta peak (which maybe broadened for nonzero $\alpha$);
 thus we expect that the Fourier transform of the
Hadamard matrix will result in a mixture of functions or measures with a range
of complexity. This is our primary motivation for studying the product 
$G_N H_N$, the Fourier transform of the Hadamard transform.

\section{The matrix elements of $G_N H_N$}

The matrix elements of $G_N H_N$ are evaluated economically as a product of
$K$ trigonometric terms. Using the matrix representations of $G_N$ and $H_N$ we 
get 
\beq
(G_N H_N)_{kn}=\f{1}{N} \sum_{l=0}^{N-1} e^{-2 \pi i (k+\alpha)(l+\alpha)/N} \, 
(-1)^{\sum_{j=0}^{K-1} b_ja_j}
\eeq
where $l=\sum_{j=0}^{K-1} 2^j b_j $, and $n=\sum_{j=0}^{K-1} 2^j a_j. $
Thus performing the independent sums over the $b_j$, and after 
some simplifications, we get
\beqa
(G_N H_N)_{kn}=&&e^{-i \pi (k+\alpha)(N-1+2 \alpha)/N}\,
 e^{-i \pi \sum_{j=0}^{K-1}a_j/2}\,\times \nonumber \\ &&\prod_{j=0}^{K-1}\cos\left[
 \f{\pi}{N}(k+\alpha)2^j + \f{\pi}{2} a_j \right]
 \eeqa
 
Thus we are led to the study of the following class of functions
which are power spectra's of the columns of the Hadamard matrix:
 $|(G_N V_n)_k|^2=|(G_N H_N)_{kn}|^2\equiv f_{n}(k)$
\beq
f_{n}(k)= \prod_{j=0}^{K-1}\cos^2\left[ \f{\pi}{N}(k+\alpha)2^j
 + \f{\pi}{2} a_j \right]
\eeq
We view these as a function of $k$ for a fixed $n=a_{K-1}a_{K-2}\cdots 
a_{0}$. They satisfy the normalization, that follows from the unitarity 
of $G_N H_N$,  
\beq
\sum_{k=0}^{N-1}f_{n}(k)=1
\eeq
and we will in fact treat $f_n(k)$ as a probability measure. We are 
interested in the limit $N\rightarrow \infty$ (or $K \rightarrow \infty$).
If there exist sequences of $n$ that lead to limiting distributions we are 
especially interested in these. In the following we use the notation that
 $(s)_m$ is an $m$-fold repetition of the binary string $s$. If $n$ is of this
 form we also denote $f_n(k)$ as $f_{(s)}(k)$. For instance the
 case when $n=N-1=(1)_K$ leads to the power spectrum of the Thue-Morse sequence
 that is well-known to limit to an multifractal measure \cite{Luck}. In this case
 \beq
f_{(1)}(k)= \prod_{j=0}^{K-1}\sin^2\left[ \f{\pi}{N}(k+\alpha)2^j \right].
\eeq
This has been particularly studied when $\alpha=0$ and found to limit to a
multifractal with a correlation dimension $D_2=0.64$ \cite{Luck}. 

\section{The participation ratio or the correlation dimension}
To probe the limiting functions, if they exist, for multifractality we
test for the scaling relation
\beq
P_n^{-1}=\sum_{k=0}^{N-1}f_{n}^2(k) \sim N ^{-D_2}.
\eeq
The left hand side of this is also interpreted as the inverse participation 
ratio, its inverse, $P_n$, being the effective spread of the power spectrum, an estimate of the 
number of ``frequencies'' ($k$) that participate in it. If $D_2$=0, the frequencies are
localized (Bragg peaks of crystallography), if $D_2$=1, it is a situation expected
 of power spectra's of random sequences. In the intermediate range are multifractals, with
  structures at many scales. 

Consider the class of functions that result as $K$ tends to $\infty$ along even numbers,
and $n=(01)_{K/2}$. Equivalently $n=(N-1)/3$, and the functions are $f_{(01)}(k)$. We also
simultaneously consider the closely related functions $f_{(10)}(k)$. Both these tend
to multifractal measures as $K \rarrow \infty$ with $D_2\approx 0.57$. The principal peaks
of $f_{(10)}(k)$ are at $1/5,4/5$, while those of $f_{(01)}(k)$ are at $2/5,3/5$. Taken together 
these peaks constitute a {\it period-4} orbit of the doubling map
 $x \mapsto 2x \, (\mbox{mod} 1)$. The peaks of the Fourier transform of the Thue-Morse
 sequence, or of $f_{(1)}(k)$ are at the period-2 orbits of the doubling map, i.e. at $1/3,2/3$.
 We see these and a few other such functions in Fig.~(1). The scaling of the 
 participation ratio of these measures and the corresponding correlation dimension
 are shown in Fig.~(2), which shows that indeed these measures are  
 multifractals. In the case of strings of the form $(001)$ for instance, $K$ is 
 taken to be multiples of $3$ and so on.
 One interesting observation from this figure is that it appears
 that the more 1 there are in the string $s$, the more is the dimension $D_2$,
 so that the power-spectrum of the Thue-Morse sequence may have the maximum possible $D_2$ 
 value in this class of multifractals. Of course the string $(0)_K$ is not a multifractal
 at all, and $D_2=0$ in this case.
 
\begin{figure}
\label{fracmeas}
\includegraphics[width=3.5in]{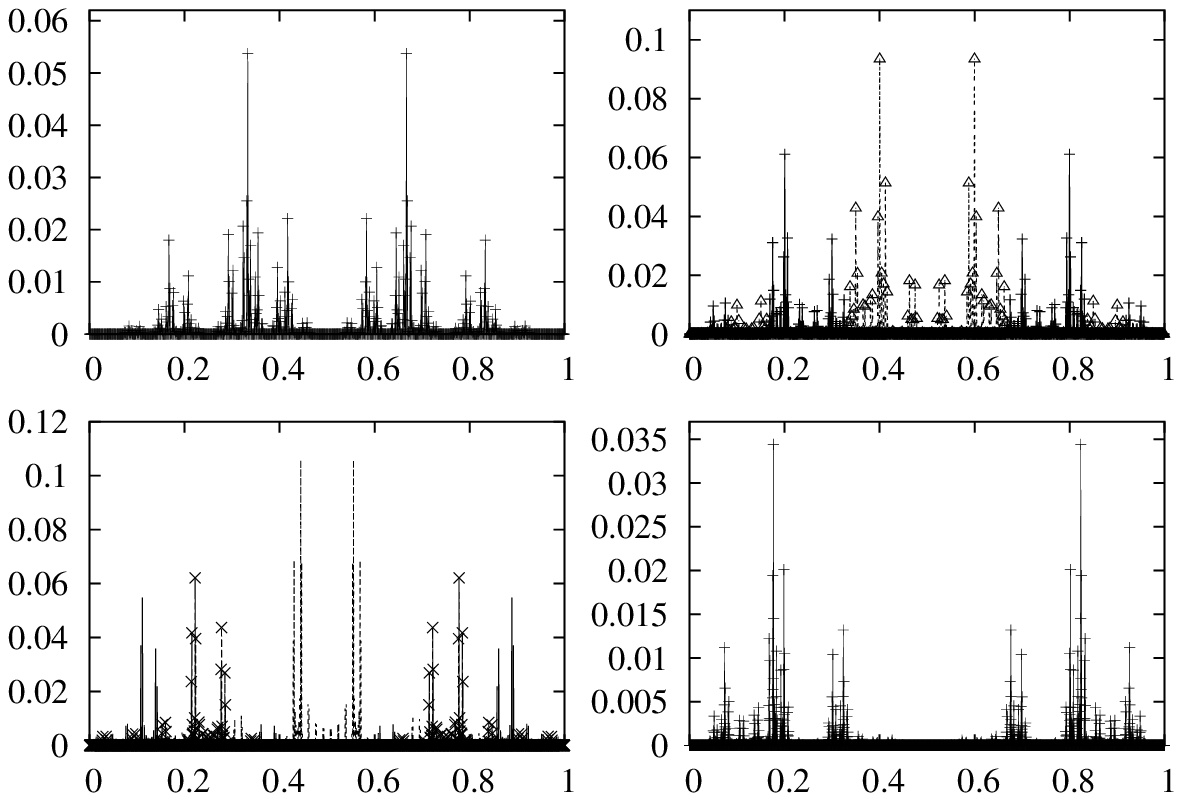}
\caption{The fractal measures $f_{(1)}$, $(f_{(01)},f_{(10)})$, $f_{(1110)}$,
$(f_{(001)}$, $f_{(010)},f_{(100)})$  are plotted clockwise. $K=10$ in the first case and $K=12$ in the
others, while $\alpha=0$ uniformly. We have grouped some functions and shown them
in different styles in the figure.}
\end{figure}

\begin{figure}
\label{d2lines}
\includegraphics[width=3.5in]{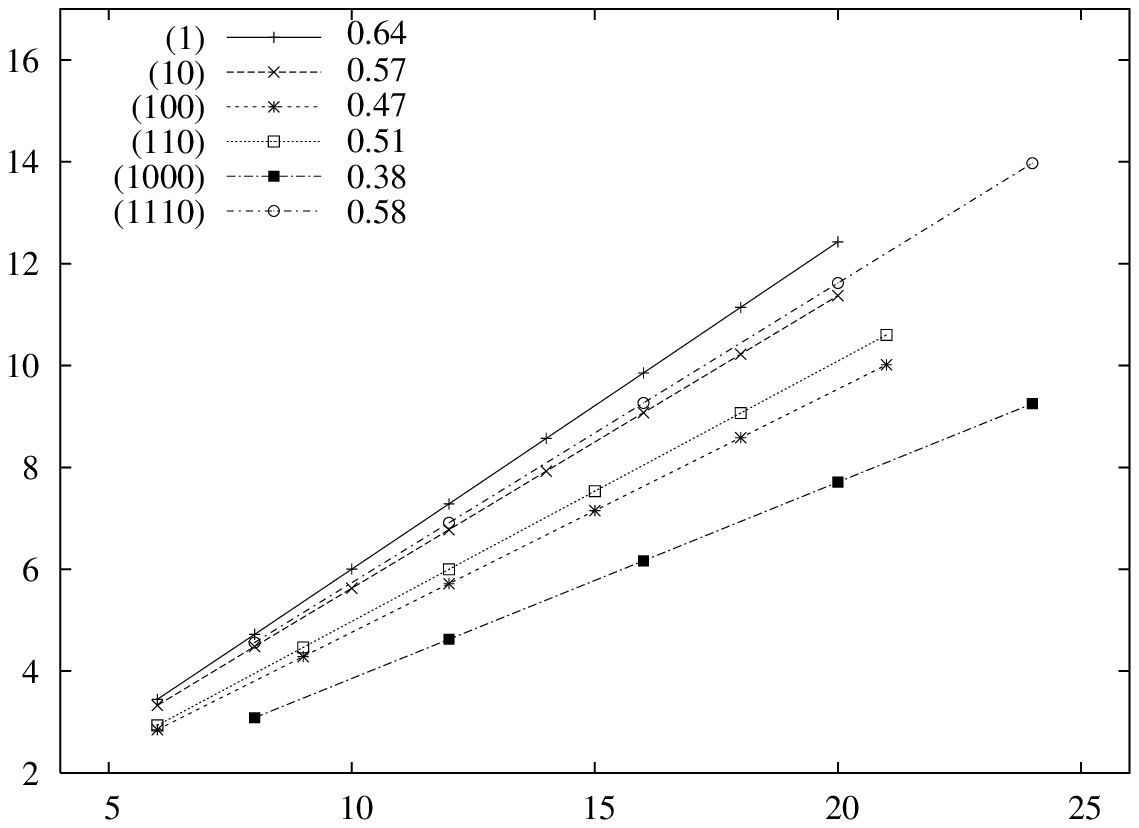}
\caption{The scaling of the participation ratio $P_n$ with $N$.  $\log_2(P_n)$ is plotted 
against $\log_2(N)=K$, and $n=(s)_m$, where the string $s$ is given
in the key to the figure, and $m$ is a whole number such that  $K/m$ is
the length of the string $s$. The slopes of the lines are also given in the key.}  
\end{figure}

For a given $N$, the participation ratio $P_n$ of the Fourier transform for the various
columns $n$, has a range of values that indicates the localization in the conjugate
basis. We show in Fig.~(3) the participation ratios for the case $N=1024$
and $\alpha=1/2$.
\begin{figure}
\label{PR}
\includegraphics[width=3.5in]{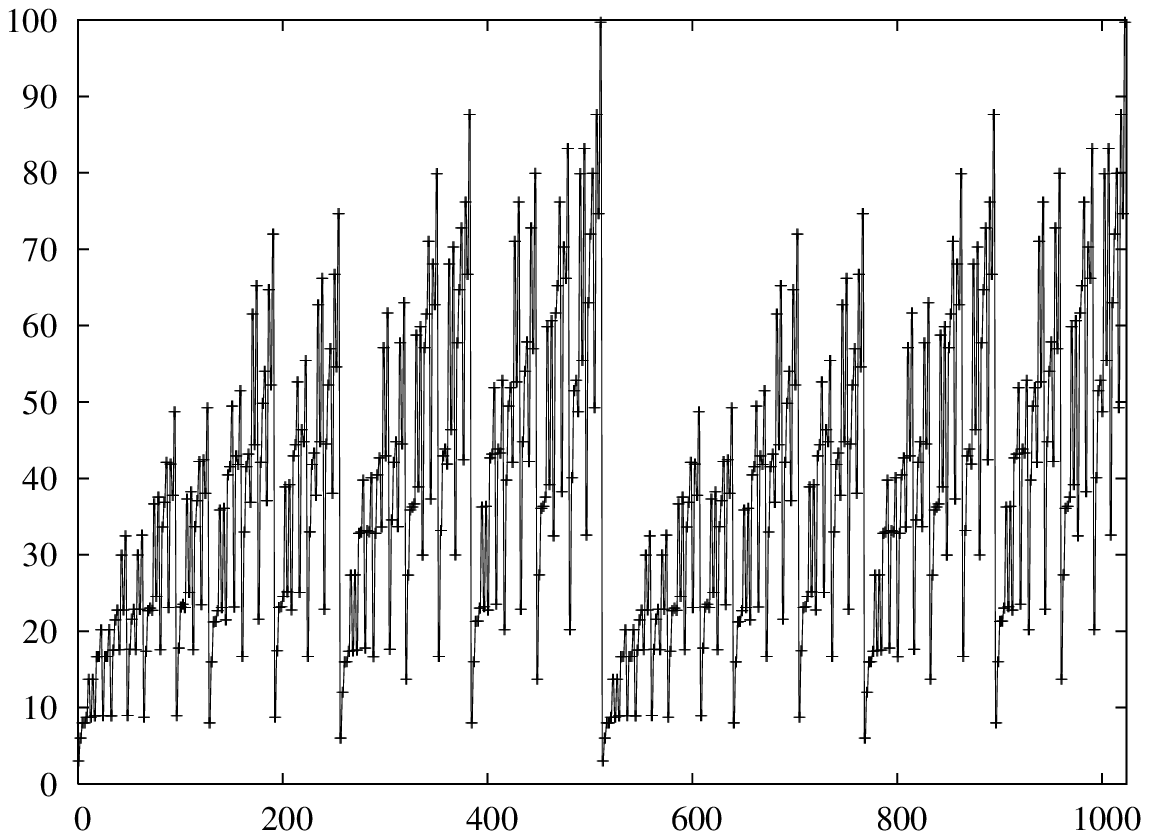}
\caption{The participation ratio $P_n$ vs $n$ for $N=1024$ and $\alpha=1/2$. }  
\end{figure}
We see here the intricate way in which the columns of the Hadamard matrix are arranged.
The largest participation ratio occurs for the last column of the Hadamard
matrix which corresponds to the Thue-Morse sequence. 
We find a similar behaviour for other values of $\alpha$, as well as other 
measures of localization such as the entropy. We include this parameter as 
for the application we have in mind, namely the quantum bakers map, 
$\alpha=1/2$ is pertinent.

  We now briefly discuss the sequence whose power spectrum is $f_{(01)}(k)$.
 Recall that the corresponding sequence for $f_{(1)}(k)$ was the Thue-Morse
sequence $\{1,-1,-1,1,-1,1,1,-1, \cdots \}$. The $n$-th term of this sequence 
is $t_n=(-1)^{\sum_{i=0}^{K-1} a_i}$ where again the $a_i$ are bits of the binary expansion
of $n$. Similarly the $n$-th term of the sequence whose power spectrum is $f_{(01)}(k)$ is 
\beq
t_n= (-1)^{\sum_{i=0,2,\ldots} a_i}.
\eeq
The first few terms of this sequence are 
$\{1,-1,1,-1, -1,1,\\-1,1, 1,-1,1,-1,-1,1,-1,1,-1,1,-1,1,
\ldots \}$. To write an concatenation rule we use the fact that this is formed by repeated
 outer product of $(1,-1,1,-1)^T$ and get  
\beq
S(k+1)=S(k) \ovl{S(k)} S(k) \ovl{S(k)}
\eeq
Where $S(k)$ is the $k$-th generation of the sequence, with $S(0)=1$, and $\ovl{S(k)}$ is the complementary set where
$1$ and $-1$ are interchanged. While we found similar rules and sequences elsewhere \cite{Sloane}, we did not
 find this exact one. It also appears that the inflation rules $ A \rightarrow ABAB, \;
  B \rightarrow BABA$ produces this sequence.

 \section{Connections to quantum chaos}
 
 So far we have introduced the measure and discussed some of their mathematical 
 properties. Here we make explicit their relevance to quantum chaos. 
The classical baker's map \cite{LL}, $T$, is the area preserving transformation
of the unit square $[0,1)\times [0,1)$ onto itself, which takes a
phase space point $(q,p)$ to $(q',p')$ where $(q'=2q,\, p'=p/2)$ if
$0\le q<1/2$ and $(q'=2q-1,\,p'=(p+1)/2)$ if $1/2\le q<1$.  The
stretching along the horizontal $q$ direction by a factor of two is
compensated exactly by a compression in the vertical $p$
direction. The repeated action of $T$ on the square leaves the phase
space mixed, this is well known to be a fully chaotic system that in a
mathematically precise sense is as random as a coin toss. The
area-preserving property makes this map a model of chaotic two-degree
of freedom Hamiltonian systems, and the Lyapunov exponent is
$\log(2)$ per iteration.  

 The baker's map was quantized by Balazs and Voros \cite{BalVor}, while 
Saraceno \cite{Saraceno} imposed anti-periodic boundary conditions, and this leads
to the quantum baker's map, in the position representation, that we use
in this Letter:
 \beq B=G_N^{-1}\left( \begin{array}{cc}G_{N/2} &0
\\0 & G_{N/2} \end{array}\right), \eeq where $(G_{N})_{mn}=\br
p_m|q_n\kt = \exp[-2 \pi i (m+1/2)(n+1/2)/N]/\sqrt{N}$.
The Hilbert space is finite dimensional, the dimensionality $N$ being the 
scaled inverse Planck constant $(N=1/h)$, where we have used that the phase-space
area is unity. The position and momentum states are denoted as $|q_n\kt$ and $|p_m\kt$,
where $m,n=0,\cdots,N-1$ and the transformation function between these bases is the
finite Fourier transform $G_N$ given above.

The choice of anti-periodic boundary conditions fully preserves parity
  symmetry, here called $R$, which is such that $R|q_{n}\kt = |q_{N-n-1}\kt.$ 
  Time-reversal symmetry is also present and implies in the context of
  the quantum baker's map that an overall phase can be chosen such that the
momentum and position representations are complex conjugates: $G_N
\phi=\phi^{*}$, if $\phi$ is an eigenstate in the position basis. 
$B$ is an unitary matrix, whose repeated application is the quantum version
 of the full left-shift of classical chaos. There is a semiclassical trace formula,
 which, based on the unstable periodic orbits, approximates eigenvalues \cite{Almeida}.

The nature of quantum chaotic eigenfunctions is intriguing as they display a bewildering
variety of patterns, that can sometimes be partially attributed to classical periodic orbits.
 This phenomenon called scarring \cite{Heller} is apparently in conflict with another observed behaviour 
namely their similarity to random matrix eigenfunctions \cite{Haake}. The quantum bakers map
affords us an opportunity to study such states in a simple setting, however no
known analytical formulae exist. We had proposed a few ansatz for a variety of states based
upon the Thue-Morse sequence and its Fourier transform \cite{MLpre}. These could sometimes reproduce
states to more than $99\%$. In this work we point to other measures such as $f_{(01)}$ that 
also play a role in the spectrum of the quantum bakers map. 

In Fig.~(4) we see two examples of states of the quantum bakers map 
in the position basis for $N=1024$, along with their Hadamard transforms. 
The Hadamard transforms are particularly simple, the Thue-Morse state, $\psi_{tm}$, being the
first \cite{MLpre}. We show it here for
comparison with another state whose Hadamard transform has distinct peaks at 
around $N/3$ and $2N/3$ implying that it is likely that the measures $f_{(01)}$ 
and $f_{(10)}$ discussed above are relevant for these states. There are 
usually more than one such state, we show here a particularly ``clean" state,
in terms of its Hadamard transform, and for the purpose of this paper call it 
$\psi_{(01)}$.
 
Indeed we can construct the ansatz
\beq 
\phi = (\gamma_1+\gamma_1^*G^{-1})\, V_{(N-1)/3}
+
(\gamma_2+\gamma_2^*G^{-1})\, V_{2(N-1)/3}
\eeq
where $N$ are powers of $4$. This includes both the columns at $(N-1)/3$ and 
$2(N-1)/3$, {\it along with their Fourier transforms}. The latter are included
due to the presence of time-reversal symmetry. In fact they dominate the state
and is the motivation for our study. This ansatz has two complex constants
$\gamma_1$ and $\gamma_2$ which we determine numerically so that its overlap
with the actual state $\psi_{(01)}$ is the maximum.
 For instance in the case $N=64$, we were able
to find $\gamma$ such that the overlap $|\br \phi|\psi_{(01)}\kt|^2 \approx 0.75$.
A comparison of the spectral measures for the Thue-Morse sequence, as well
as $f_{(01)}$ and $f_{(10)}$ in Fig.~(1) with the actual eigenfunctions in Fig.~(4)
 show the similarity between them. Notice that the actual wavefunction is more
 closely related to the Fourier transforms of a linear combination of the
 columns of the Hadamard matrix. The peaks of this wavefunction can be identified with
 the period-4 orbit starting from $(q=1/5,p=4/5)$, and has been noted earlier
 the $q$ part of this corresponds to the peaks in the measures $f_{(01)}$ and 
 $f_{(10)}$.
 
\begin{figure}
\label{efns}
\includegraphics[width=3.5in]{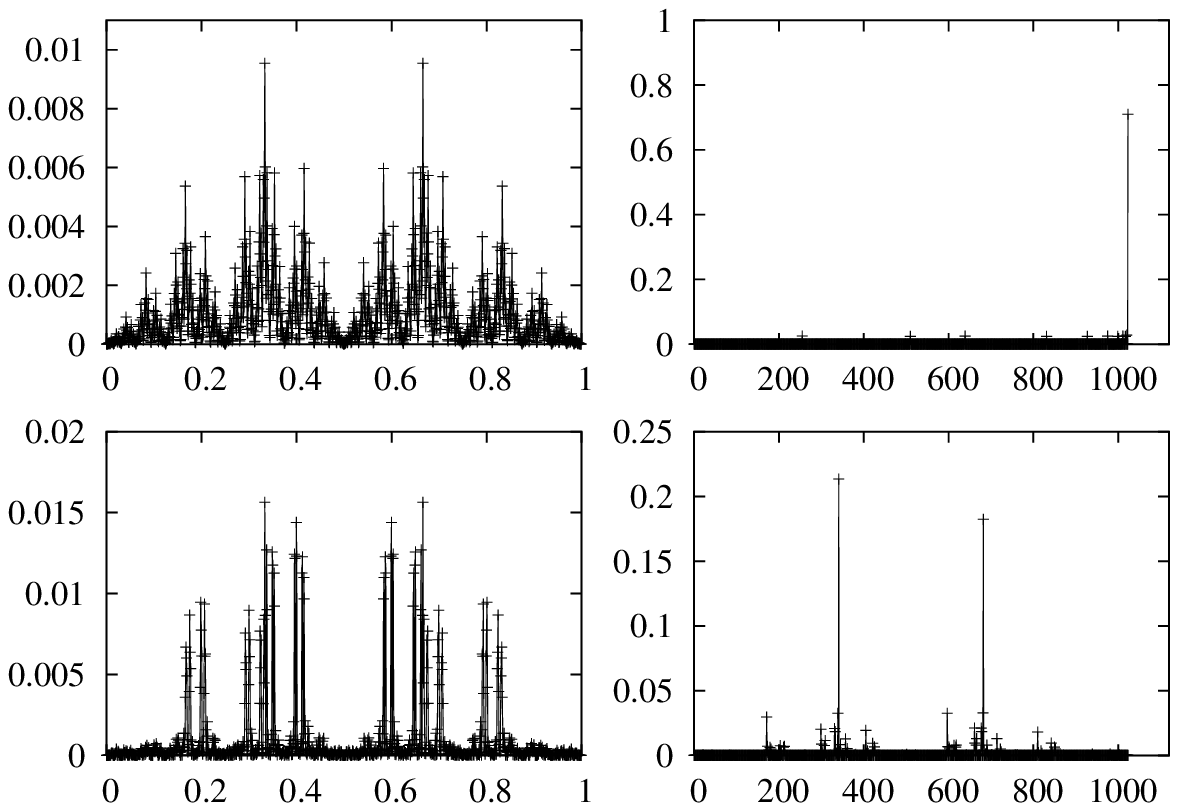}
\caption{The Thue-Morse state $\psi_{tm}$ (top left) and a state called in the text $\psi_{(01)}$ 
(bottom left),
along with their corresponding Hadamard transforms (right). In this case $N=1024$ and 
the peaks of the Hadamard transform of $\psi_{(01)}$ occur at $(N-1)/3$ and $2(N-1)/3$.} 
\end{figure}

We have concentrated on convergent measures as $n=(s)_m$ tends to infinity and 
$s$ is a bit string repeated $m$ times. We can have other sequences of $n$ that lead
to convergent measures, in particular those whose binary expansions ends in a string 
of $1$ or $0$. In the former case we have already used such measures to describe
a family of states of the bakers map that are scarred by period-2, period-4
and associated homoclinic orbits \cite{MLpre}. As spectral measures they are different depending
on the finite string that precedes the infinite string of 1; however their correlation
dimension $D_2$ seems to be the same as that of the spectrum of the Thue-Morse sequence,
namely $0.64$. As we noted earlier this seems to be consistent with our view that more
the 1, more the dimension $D_2$. If the string $a_i$ is itself not periodic, say when
$n$ is tending to infinity such that $n/2^K$ tends to an irrational number, there does
not seem to be any convergent measures.

In summary, we have introduced a class of simple functions that limit to multifractals,
have interesting connections to sequences and to a simple model of quantum chaos. We have 
done this by combining two well-known, well-studied and standard transforms, namely the
Fourier and the Hadamard.

\bibliographystyle{ncnsd}


%

\end{document}